\documentclass[sigconf]{acmart}
\AtBeginDocument{%
  \providecommand\BibTeX{{%
    \normalfont B\kern-0.5em{\scshape i\kern-0.25em b}\kern-0.8em\TeX}}}

\copyrightyear{2023}
\acmYear{2023}
\setcopyright{rightsretained}
\acmConference[CPS-IoT Week Workshops '23]{Cyber-Physical Systems and Internet of Things Week 2023}{May 9--12, 2023}{San Antonio, TX, USA}
\acmBooktitle{Cyber-Physical Systems and Internet of Things Week 2023 (CPS-IoT Week Workshops '23), May 9--12, 2023, San Antonio, TX, USA}
\acmDOI{10.1145/3576914.3587491}
\acmISBN{979-8-4007-0049-1/23/05}

\begin{document}

\title{Towards Usable Parental Control for Voice Assistants}


\author{Peiyi Yang}
 \affiliation{
 \institution{University of Virginia}
 \country{USA}
 }
\author{Jie Fan}
 \affiliation{
 \institution{University of Virginia}
 \country{USA}
 }
\author{Zice Wei}
 \affiliation{
 \institution{University of Virginia}
 \country{USA}
 }
\author{Haoqian Li}
 \affiliation{
 \institution{University of Virginia}
 \country{USA}
 }
\author{Tu Le}
 \affiliation{
 \institution{University of Virginia}
 \country{USA}
 }
 \additionalaffiliation{University of California, Los Angeles, USA}
\author{Yuan Tian}
 \affiliation{
 \institution{University of California, Los Angeles}
 \country{USA}
 }
 \additionalaffiliation{University of Virginia}


\begin{abstract}
Voice Personal Assistants (VPA) have become a common household appliance. As one of the leading platforms for VPA technology, Amazon created Alexa and designed Amazon Kids for children to safely enjoy the rich functionalities of VPA and for parents to monitor their kids' activities through the Parent Dashboard. Although this ecosystem is in place, the usage of Parent Dashboard is not yet popularized among parents. In this paper, we conduct a parent survey to find out what they like and dislike about the current parental control features. We find that parents need more visuals about their children's activity, easier access to security features for their children, and a better user interface. Based on the insights from our survey, we present a new design for the Parent Dashboard considering the parents' expectations.
\end{abstract}

\ccsdesc[500]{Security and privacy~Domain-specific security and privacy architectures}
\ccsdesc[500]{Security and privacy~Usability in security and privacy}

\keywords{Voice Assistant, Smart Speaker, Security, Privacy, Safety, Parental Control, User Interface, Amazon Alexa, Alexa Kids}

\maketitle

\section{Introduction}
The development of Internet of Things (IoT) technology is growing rapidly and competition for the control hub of those IoT devices has heated up over the last few years. There are some forerunners in the business of providing voice personal assistant (VPA) services, such as Amazon Alexa and Google Home.

While VPAs can make our daily routines automated, there are some serious risks from using VPAs that we can not ignore, especially for households that have children. The voice authentication systems on commercial VPAs are found to be not robust~\cite{yuan2018echoattack}. Hence, underage children could easily gain access to content that is for adult audiences, which may contain inappropriate language content. Amazon Alexa has been the most popular VPA platform. Researchers have found a lot of issues with the Alexa platform and its applications (hereinafter referred to as ``skills'') in the past~\cite{kumar2018squatting, zhang2019dangerousskills, edu2021skillvet, liao2020measuring, le2022skillbot}. Even with the dedicated Alexa kid skills with a stricter vetting process, children users are still vulnerable to private data collection and inappropriate content risks from such skills~\cite{le2022skillbot}. 

Luckily, Amazon Kids provides parents with an Alexa Parent Dashboard to monitor their children's activities. The Parent Dashboard also has security features to protect the kids. However, it is unclear whether the current design of the Parent Dashboard is effective and whether the security features such as Alexa skills allowlisting are useful to parents. Although the Parent Dashboard provides basic functionalities, there might still be room for improvements to attract more users and ensure children's safety. 

In this work, we aim to understand what limitations the current parental control scheme of VPA has and how we can improve the design to improve user experience. Our aim is to build a new Parent Dashboard for parents that can replace Alexa Parent Dashboard.

This work contributes the following aspects:
\begin{itemize}
    \item First, we design and conduct an online survey with Alexa users who have kids at home to understand what parents expect from Amazon Kids and how the current parental control scheme aligns with their expectations.
    \item Second, we also study the parents' awareness of the security and privacy risks introduced by Alexa and what actions they take to mitigate these risks.
    \item Finally, using the insights from our survey, we design a new Parent Dashboard for VPA to improve the parental control user experience while ensuring children's safety.
\end{itemize}
\section{Related work}
Previous research has investigated children's safety on the Internet and compliance. Several user studies were conducted to provide insights into how parents and children understand the technology and make privacy decision~\cite{mertala2019young, cranor2014parents}. Besides, many analyses of mobile apps identified violations regarding data collection and non-compliant privacy policies~\cite{zimmeck2016automated, reyes2018won}. Websites made for children users were also found to implement hidden tracking or malicious algorithms~\cite{vlajic2018online, cai2013online}. Other studies investigated inappropriate content and privacy issues in child-directed voice applications~\cite{le2022skillbot} and smart toys~\cite{manches2015three, mcreynolds2017toys, chu2018security, mahmoud2018towards, streiff2018s, valente2017security}, showing children are vulnerable to malicious content and parents are actually concerned about such risks. Addressing such issues, several frameworks and recommendations were proposed for ensuring children's safety~\cite{rafferty2017towards, haynes2017framework}. Different from these previous studies, we investigate the usability of parental control for voice assistants and propose an improved design based on the insights given by our participants.
\section{Parent Survey}
In this section, we discuss our recruitment strategy, ethical considerations, how we design our survey, and the results.

\subsection{Recruitment}

We recruited 140 participants on Prolific to participate in our study. Participants were required to be fluent in English with at least one child and have at least one Amazon Echo device. Our survey consisted of 33 questions, and the payment was \$9.54/hr rate suggested by Prolific for completing our survey.


\subsection{Survey Pretest}
Pilot study is a common practice to identify design issues and biases for surveys, including priming or confusing wording before deployment~\cite{presser2004survey}. We ran a pilot study with 7 participants to obtain feedback for our survey design and payment logistics. As a result, we improved the wording and presentation of our survey questions. Our study results reported in this paper did not include the data collected from the pilot study.

\subsection{Ethical Considerations}
Our study protocol was approved by our Institutional Review Board (IRB). The participants were asked to read our consent form carefully and sign it to participate in the study. Participation in our study was voluntary and anonymous. We did not collect any personally identifiable information. Payments were handled through Prolific and adhered to Prolific's policies.

\subsection{Survey Design and Results}
\subsubsection{Demographic}
We asked the participants to report their age, gender, and comfort level with computing technology. We allowed the ``Prefer not to answer'' option for these questions. The majority of our participants were in the ranges of 25--44 years old (66.42\% of participants) and 45--64 years old (23.36\% of participants). We also had a relatively gender-balanced participant pool as 53.28\% of participants are male and 46.72\% are female.

\subsubsection{Overview experiences on Amazon Echo Device}
We asked the participants about how frequently they used Amazon Echo devices, whether they had ever used parental control, and whether they were satisfied with the parental control mode on Alex in terms of filtering inappropriate/adult content, preventing privacy invasion, entertaining their kids, keeping track of there kids' activates, keep control over their kids' activities, and overall experience.

\begin{figure}[htbp]
\begin{center}
\includegraphics[width=\linewidth]{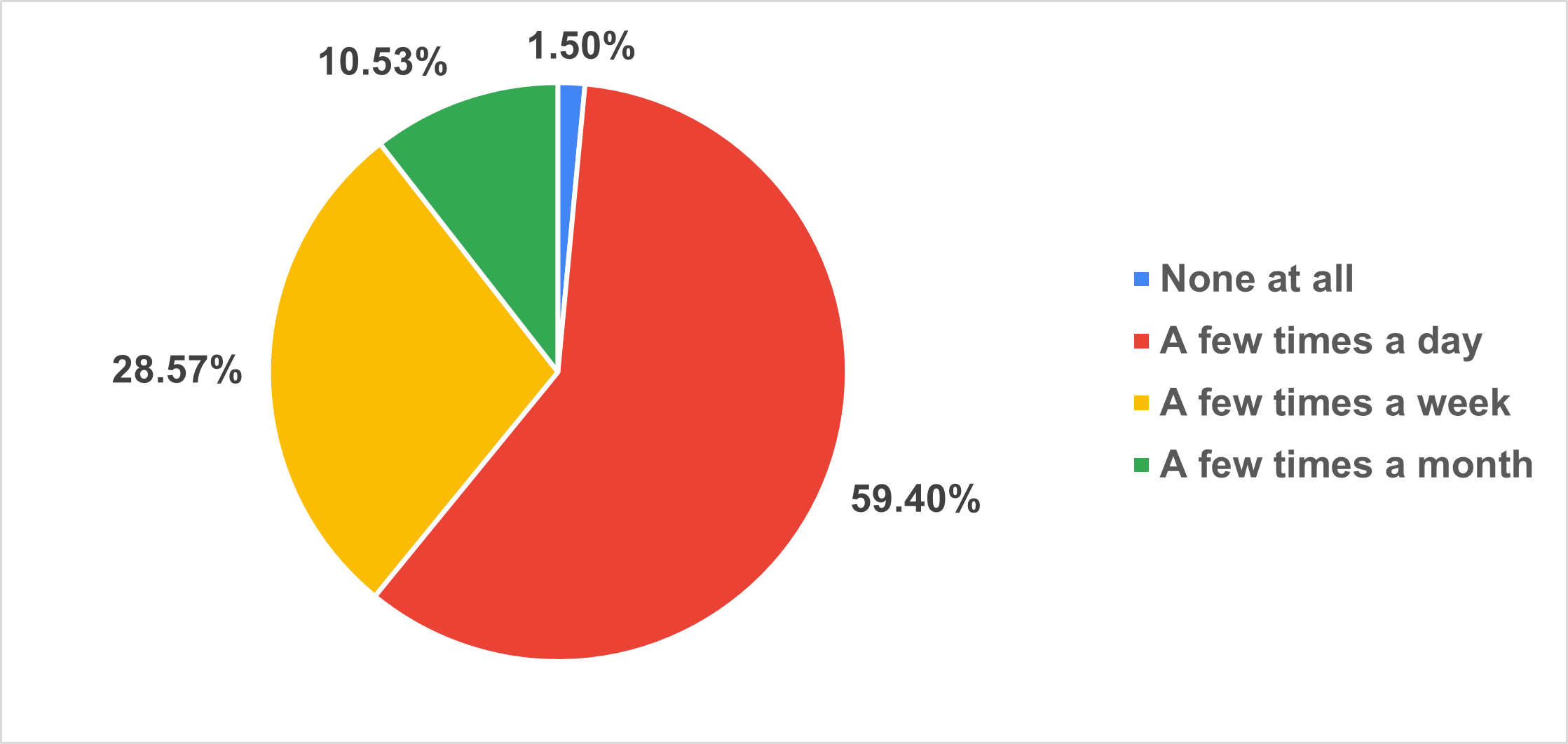} 
\caption{Participants' responses to how frequently they used Amazon Echo devices. The majority of participants used their Echo devices everyday.}
\label{fig:daily_basis}
\end{center}
\end{figure}

\begin{figure}[htbp]
\begin{center}
\includegraphics[width=\linewidth]{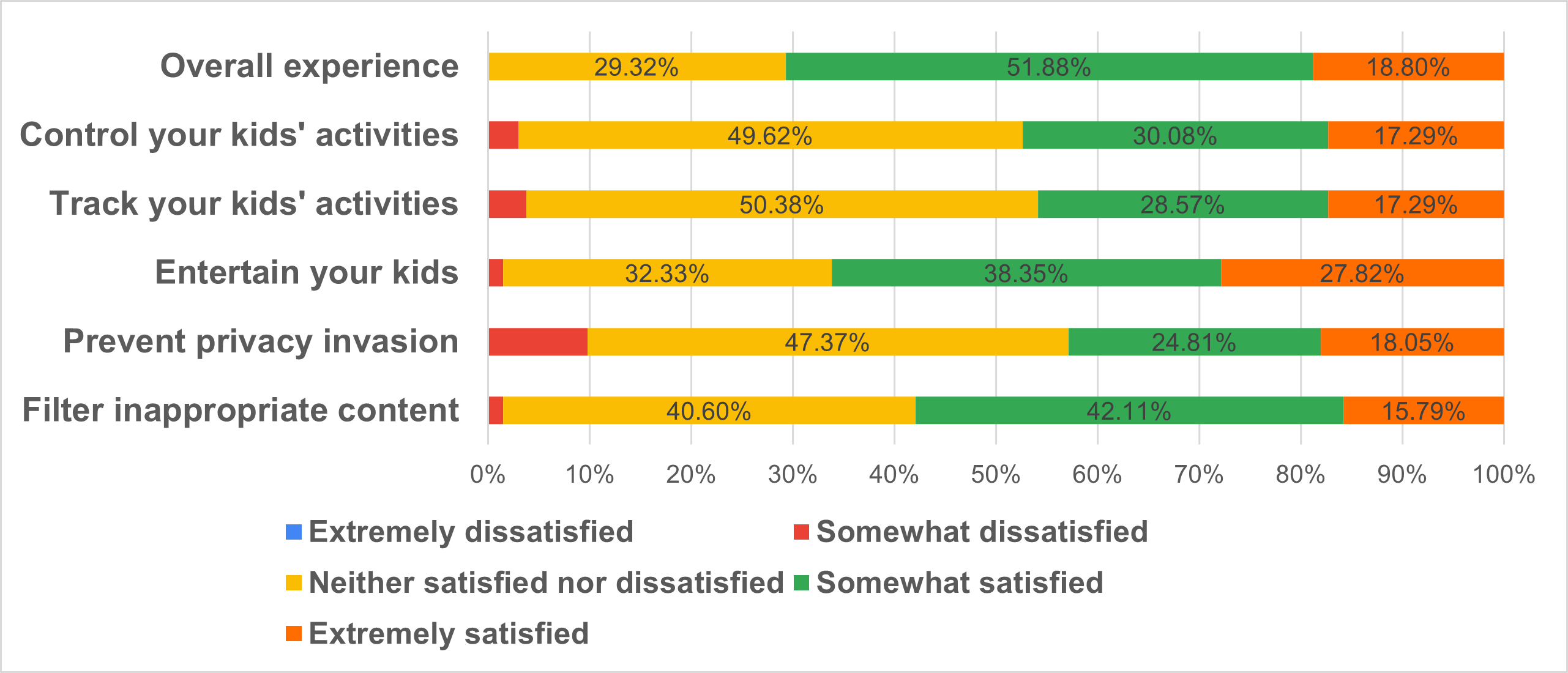} 
\caption{Participants' responses to their level of satisfaction of the contents under parental control mode on Alexa}
\label{fig:filtering}
\end{center}
\end{figure}

Our results show that around 60\% of participants used Amazon Echo devices on a daily basis (Figure~\ref{fig:daily_basis}). 50\% of the participants used parental control. 35\% did not use it since they did not know about these features. Figure~\ref{fig:filtering} shows the level of satisfaction our participants had towards parental control mode on Alexa. Overall, around 70\% were satisfied with the contents under Alexa parental control mode such as filtering inappropriate/adult content, preventing privacy invasion, entertaining their kid, keeping track of their kids' activities, and keeping control over their kids' activities. However, particularly for preventing privacy invasion, 10\% of participants were dissatisfied, which is noticeably more than the other contents. We then asked whether they thought Amazon Echo devices could potentially leak private information or have a bad influence on kids. 43\% of participants did not think it was possible for Alexa to leak private information. 72\% of participants did not believe Alexa could have a bad influence on their kids.

\subsubsection{User Interface of Parent Dashboard}
\begin{figure}[htbp]
\begin{center}
\includegraphics[width=\linewidth]{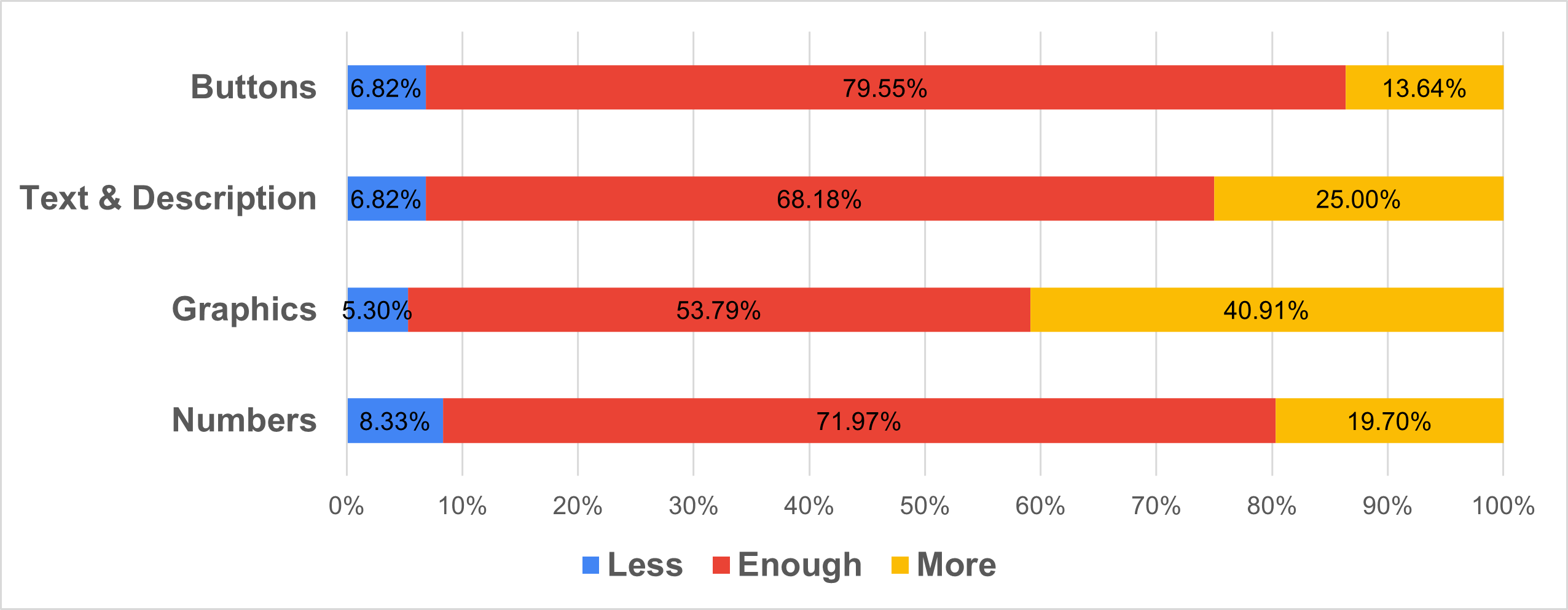} 
\caption{Participants' responses to whether they think the user interface should contain more or less numbers/graphics/buttons/text and description. Many participants (about 40\%) preferred more graphics.}
\label{fig:UI_question}
\end{center}
\end{figure}

We provided an example of how to get access to the kid's activity summary. 74\% of participants answered easy and 18\% of them thought it was neither easy nor hard. Only 4.5\% thought it was hard to find the activity summary section. Next, we asked the participants whether the user interface should contain numbers/graphics/text and description/buttons (Figure~\ref{fig:UI_question}). 70\% of the participants preferred keeping the current design for numbers. 40\% thought that the interface should include more graphics. 68\% of participants preferred the current design of texts/descriptions and buttons. We then asked them whether they would like to see more bar charts, pie charts, and calendar views showing the kid's usage. Over 95\% of participants would like to see a pie chart for daily usage. Over 70\% were interested in seeing a bar chart showing daily usage in a week. For the calendar review, the darker color squares mean more activities, and lighter color squares mean fewer activities. We found that the calendar component idea did not appeal to most of the participants (66\%). Only 34\% said they would like to see a calendar. Therefore, a calendar component might not be necessary.

\subsubsection{Parents' Security and Privacy Behaviors}
\begin{figure}[htbp]
\begin{center}
\includegraphics[width=\linewidth]{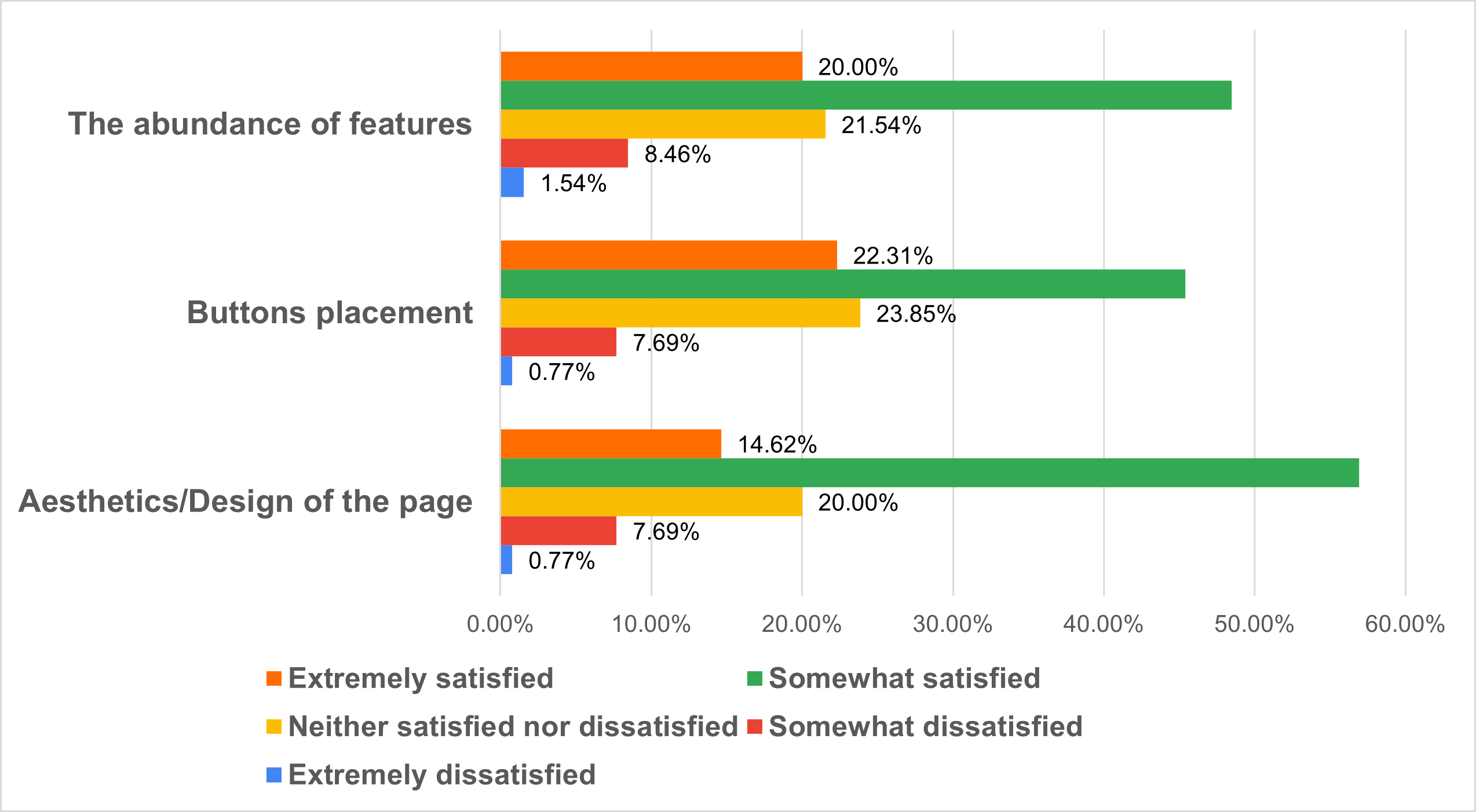} 
\caption{Participants' responses to their level of satisfaction with Alexa parental control settings. There is still room for improvement to the settings.}
\label{fig:parental_control_settings}
\end{center}
\end{figure}

We asked our participants whether it was difficult for them to enable Amazon Kids. Most participants thought it was very easy for them to access and enable Amazon Kids. Then, we asked our participants about their satisfaction with the parental control settings in the Alexa application in terms of aesthetics/design of the page, button placement, and the abundance of features (Figure~\ref{fig:parental_control_settings}). Our participants thought the design, button placement, and abundance of features were good. However, they were not extremely satisfied with it. Therefore, there is still room to revamp the aesthetics and the button placement and add more features. We further asked if the parental dashboard should contain more numbers/graphics/text/buttons. Many users wanted more graphics and descriptions of their child's activity summary. Therefore, it is important to include more graphics and descriptions for the activity summary feature. At the moment, the Alexa companion app and Alexa parental dashboard are two different apps. Thus, we asked the user whether they would like to see them combined in one single app. The majority of the users would like these two features to be combined. This result suggests that central access within a single Alexa app is better.

Next, we asked our participants about the three security features in Amazon Kids that restrict kids' behaviors when using Alexa: Add Content, Explicit Filter, and Web Access Restrictions. 

\begin{figure}[htbp]
    \centering
    \includegraphics[width=\linewidth]{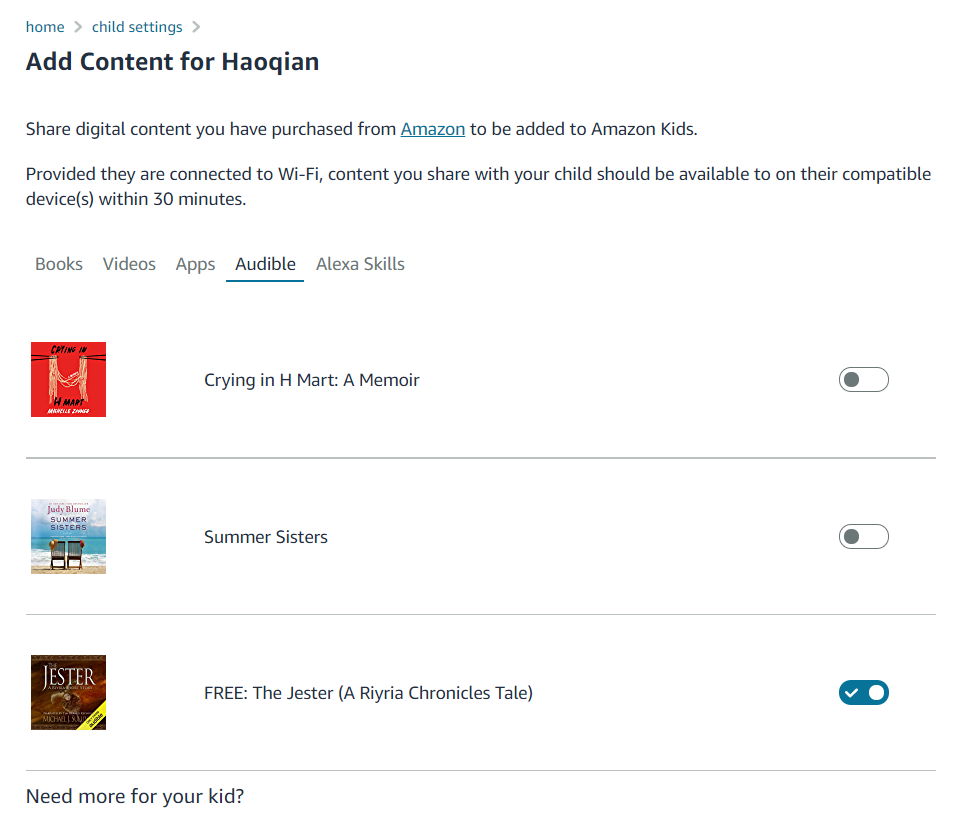}
    \caption{Add Content allows parents to make additional content available to their kids.}
    \label{fig:add-content}
\end{figure}

\begin{figure}[htbp]
    \centering
    \includegraphics[width=\linewidth]{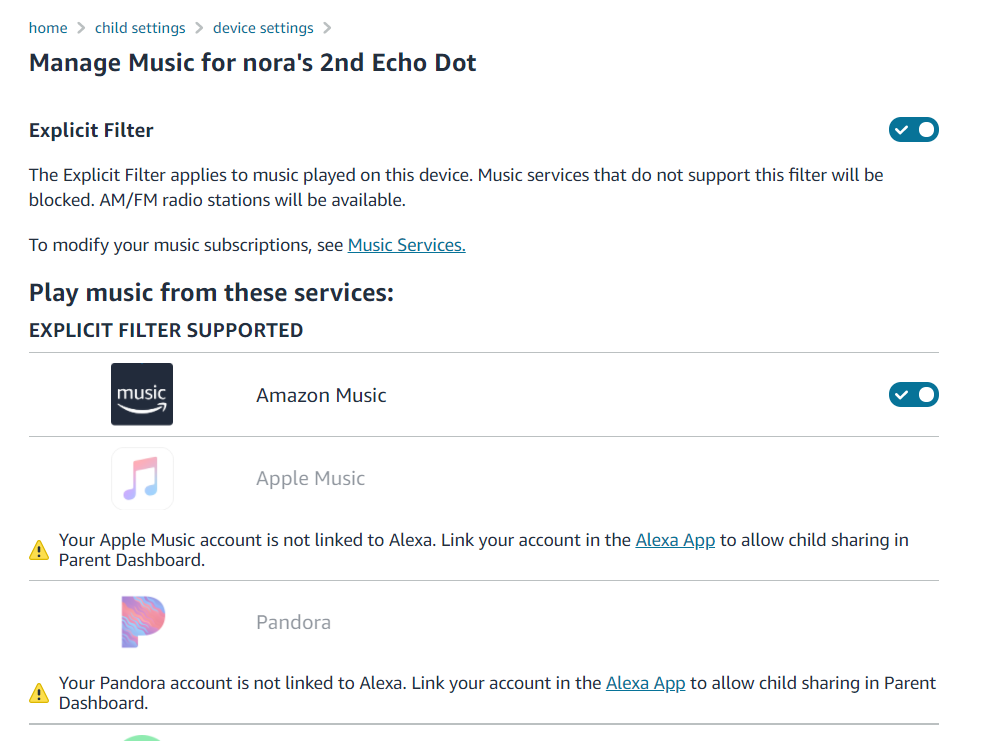}
    \caption{Explicit Filter can be turned on to protect children from explicit languages and music.}
    \label{fig:explicit-filter}
\end{figure}

\begin{figure}[htbp]
    \centering
    \includegraphics[width=\linewidth]{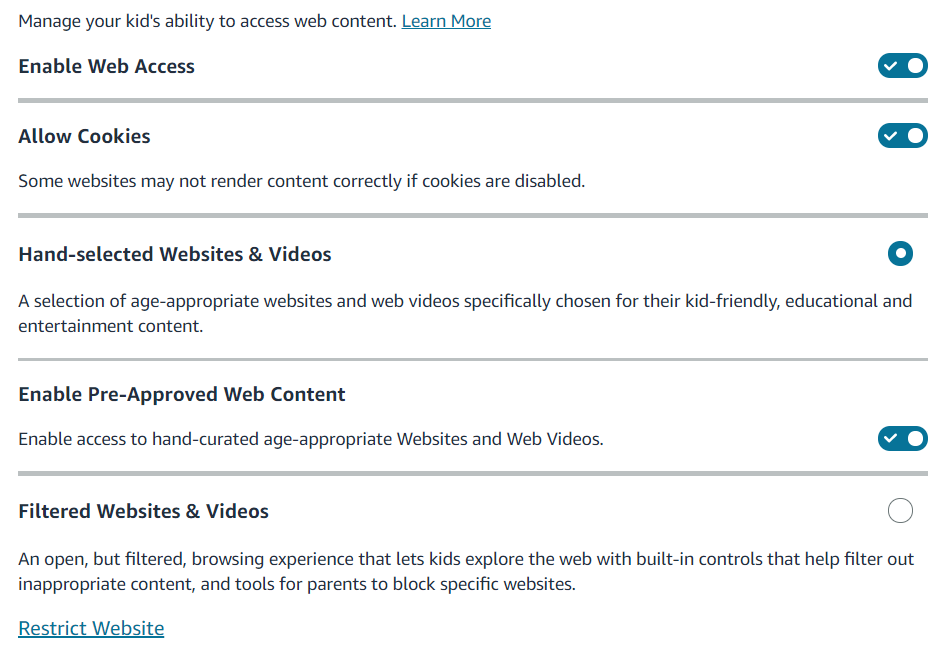}
    \caption{3 types of built-in Web Access Restrictions are provided by Amazon.}
    \label{fig:web-restrictions}
\end{figure}

As shown in Figure~\ref{fig:add-content}, Add Content allows parents to make additional Alexa Skills, Audible Books, and Kindle Books available to their kids. These additional contents are usually reviewed and trusted by the parents. We showed the participants the location of the feature and asked if they used this feature. Many participants (65\%) did not use this feature. Only 30\% of the participants used it. Out of these participants, 75\% believed this feature was effective in protecting the privacy and security of their kids. We further asked them what actions they took when their kids used Alexa Skills, Audible books, and Kindle Books. We manually checked the answers and found that almost half (47\%) indicated that no action was taken. We identified three main reasons: (1) participants believed their kids would be mature enough to not get hurt, (2) they simply were not concerned, and (3) they never thought about it until taking this survey. 19\% answered that they only allowed their kids to use Alexa when they were present. 16\% answered that they paid attention to the content their kids could have access to by manually checking their downloads. 17\% answered that their kids were not allowed to have access to these types of content at all. 
The next security feature, as shown in Figure~\ref{fig:explicit-filter}, is the explicit language filter for filtering out inappropriate music. 20\% of participants indicated their kids did not listen to music on Alexa devices. 39\% of participants said they enabled the filter, and 78\% of them believed the filter was effective. Still, 39\% of participants reported listening to music using Alexa devices without the filter. When we asked what other actions they took to protect their kids, more than half (63\%) answered they did not take any actions. The reasons roughly fall into 3 categories: (1) their kids were old enough, (2) their kids were too young, (3) there was nothing to be worried about. 30\% of the participants said that they provided direct supervision to their kids by being present, and 6\% answered that they controlled the content their kids could have access to by manually checking downloads.
The last security feature, Web Access Restrictions, as shown in Figure \ref{fig:web-restrictions}, is specifically for Alexa devices that come with a screen. With a screen, kids can use a browser to access the web and parents might want to put a limit on web access. Amazon provides three built-in types of web access restrictions: Pre-selected content only, filtered content with a URL blocklist, and filtered content without the blocklist. 50\% of participants indicated their kids had access to the web, and the number of participants for each type of restriction method was roughly the same: 31\% allowed Amazon pre-selected content, 37\% allowed filtered content with a blocklist, and 31\% allowed filtered content without a blocklist. Overall, 70\% of the participants were happy with their choice of method. 

\subsubsection{Final Remarks}
Finally, we asked participants whether they felt they had enough control by using Amazon Kids. The majority (90\%) of participants answered yes. 
Among those who were not happy with the current design, one participant explained that they did not always agree with Amazon on the definition of child-appropriate content and suggested that Amazon should invite parents from different backgrounds to share their opinions on child safety. Two participants complained there was too much control and that Alexa had gotten in the way of their parenting. As a side note, two other participants complained it was difficult to navigate the Alexa app and find what they need. One participant asked for a new feature of screen recording their kids' web activities.
\section{Improving Parental Control Dashboard Design}
In this section, we summarize the key takeaways from our survey and present our proposed design for the Parent Dashboard.

\subsection{Takeaways from Parent Survey}
Based on the insights from our survey, we propose to design a new user interface for the Parent Dashboard. In particular, the important improvements are a straightforward user interface, visuals to represent a child's activity summary, and easy access to many security features to protect children.

From the aesthetics perspective, the problem is that there is a lack of graphics and descriptions for the children's activities. The current design used bar charts to show the activities of a child in a week's time. This way of displaying activity information lacks the percentage view of pie charts and also does not show the frequency of usage. 95\% of our participants would like to see pie charts added to the Parent Dashboard. The bar charts still display valuable information, which is why 70\% decided to keep them. However, the calendar view of frequency charts is considered unnecessary by the participants. Therefore, we propose a design for our Parent Dashboard that incorporates pie charts and bar charts to visualize children's activity summaries. 

Currently, the Alexa companion app and Parent Dashboard are separate apps. Many parents preferred having them combined into a single application. Additionally, many parents did not know where to find the security features. Therefore, it is important to give easy access to the features on the Parent Dashboard. 

\subsection{Parent Dashboard Prototype Design}
We use Figma to design our prototype because it is a fast user interface prototyping platform that provides the most design functionalities. This helps make our design more professional right out of the box.
\begin{figure}[htbp]
    \centering
    \includegraphics[width=\linewidth]{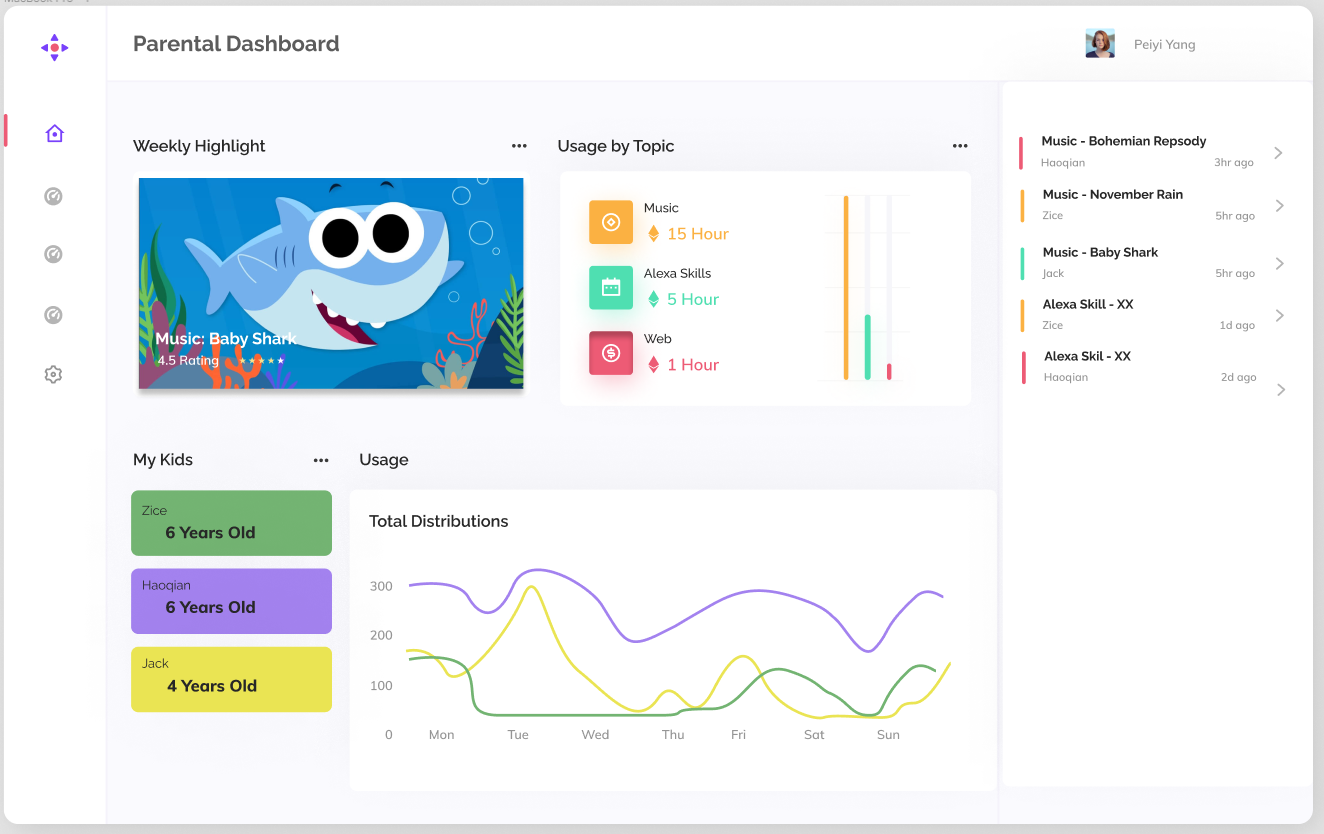}
    \caption{Parent Dashboard Homepage}
    \label{fig:web1}
\end{figure}
We have three different pages designed for the prototype. Our design philosophy is to keep what the users like about the Alexa Parent Dashboard and change/insert features the user would like to have. The first page is the Parent Dashboard homepage. We can see the new web prototype is more aesthetically designed with bigger figures, colorful charts/graphs, and a more modern design. We have changed how users view their children's activities. In Alexa Parent Dashboard, the homepage displays a list of children with hyperlinks to view their child's activity. Our prototype displays all children's activities into a line chart, so the parents can compare how much time each child has used their Alexa and see alerting information if one has over-used technology. We also include a bar chart that displays combined usage in each category. This provides parents with information about which category is used the most and the least. Parents are able to view the features (e.g., song, video, or book) that are used most frequently by their child in highlight, allowing them to know what their kid is doing at a single glance. Also, on the right-hand side, we provide the whole list of features used by their kids so that parents can quickly learn what their kids did with Alexa. By clicking on a child's info box, the parents can navigate to a particular child's activity summary page.
\begin{figure}[htbp]
    \centering
    \includegraphics[width=\linewidth]{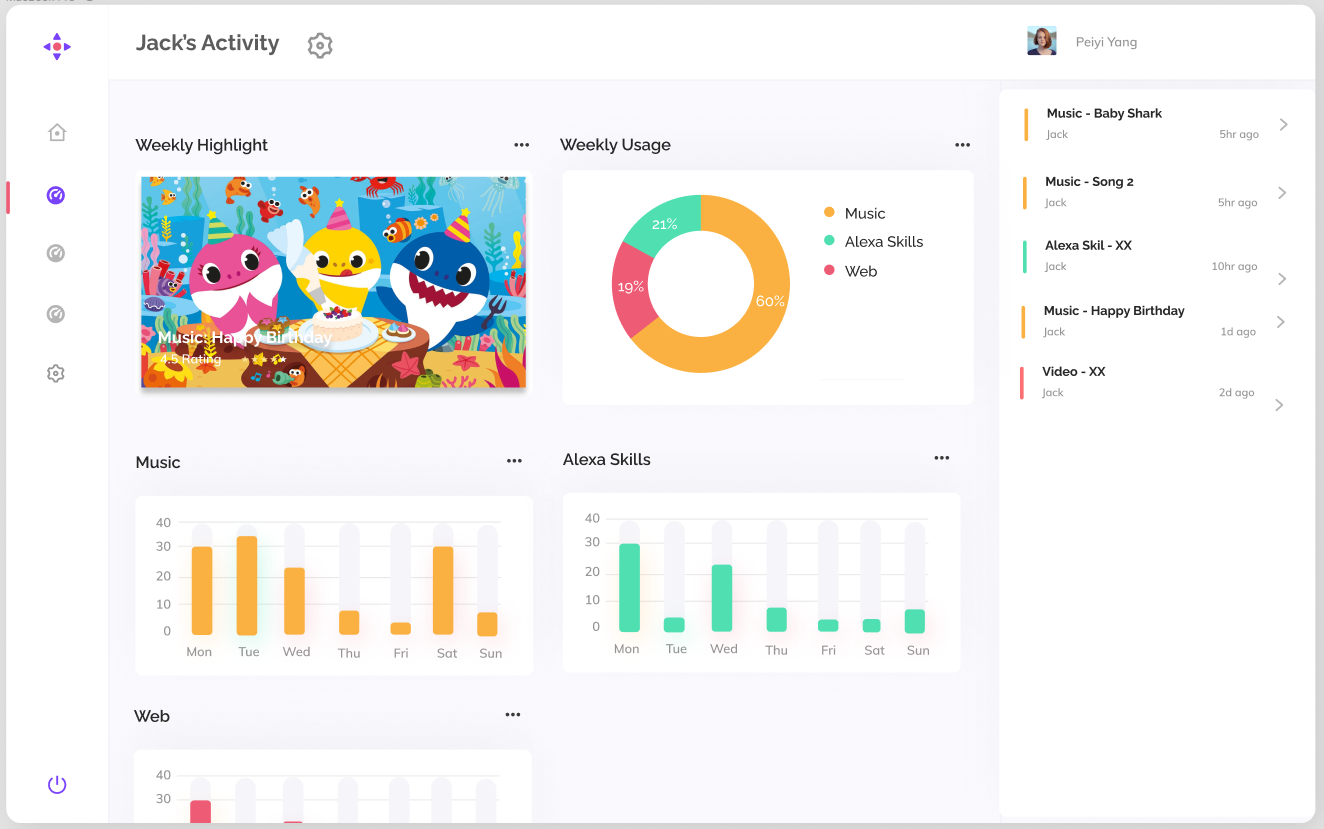}
    \caption{Single Child Activity Page}
    \label{fig:web2}
\end{figure}
The second page of our prototype is a single child's activity summary page. This page aims to show the summary and trends of a single child. It uses a similar style to the homepage and uses similar navigation logic as Alexa Parent Dashboard. On the page, we include the weekly highlight and a list of past feature usage for a single child. A new feature is a bar chart that shows the percentage of usage for a particular feature. This pie chart provides a better understanding of which feature is used the most by this child. For example, if Jack used Alexa mostly for music, his parents can learn this information at a glance and check if Jack listened to inappropriate music. 
\begin{figure}[htbp]
    \centering
    \includegraphics[width=\linewidth]{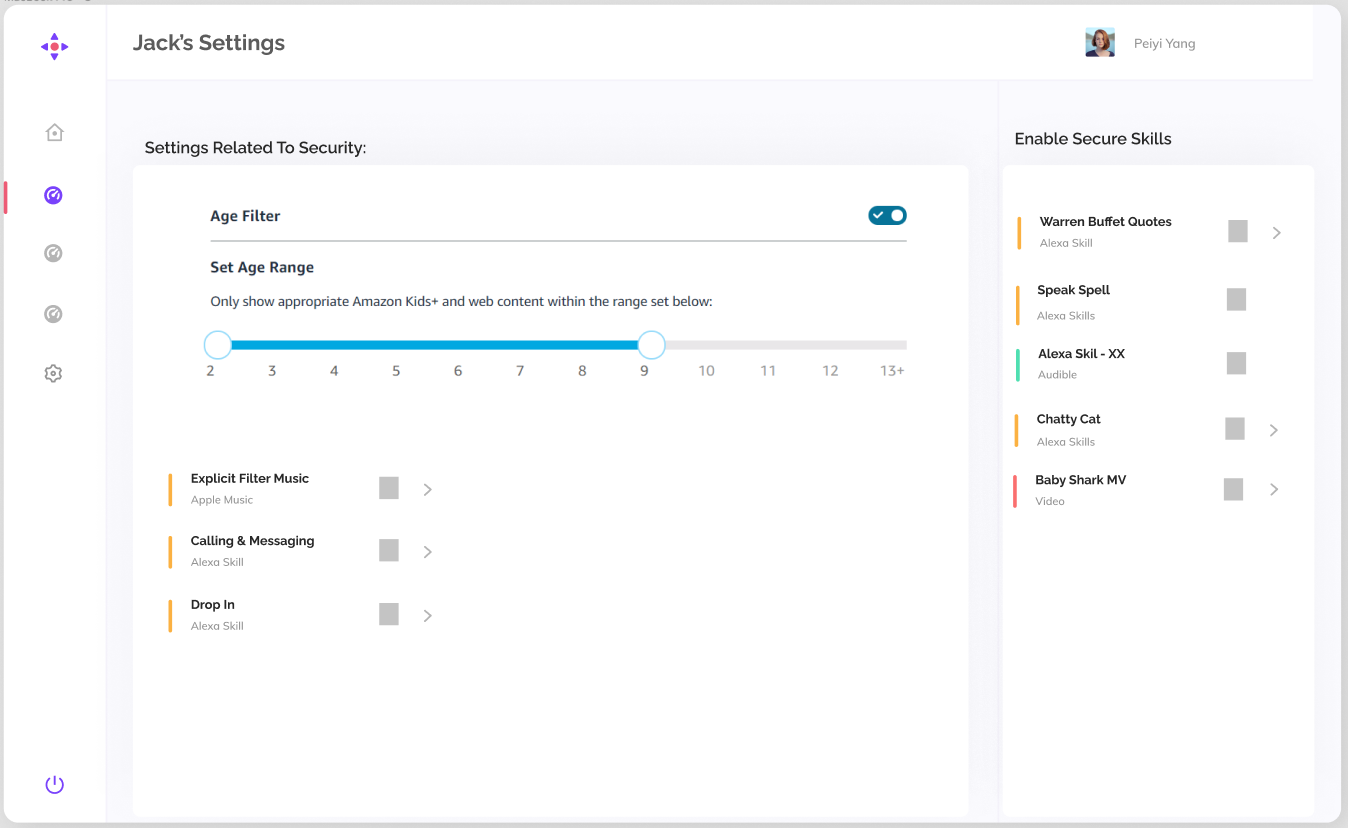}
    \caption{Single Child Setting Page}
    \label{fig:web3}
\end{figure}
The third page shown in Figure ~\ref{fig:web3} is the setting page for a single child. We decided to include the settings page because the users would like to use the security features provided by Alexa Parent Dashboard. The Alexa Parent Dashboard has amazing security features, but those features are not easily accessible. Therefore, we create a single page for the security settings. This setting page is a light version. We do not include all possible settings because we want to emphasize the security settings. 

For safe content, we include the following features: age filter, music explicit filter, calling/messaging, and parent drop-in. These features will be enabled using check-boxes as shown on the left-hand side of Figure~\ref{fig:web3}. Alexa Parent Dashboard also has a whitelist feature that our participants liked. We include this white list feature on the right-hand side of the settings page. This white list allows users to enable/disable all of their existing Alexa skills, audibles, videos, etc.

To make our front-end prototype more convincing, we add transitions between the pages. The user is able to navigate between the homepage, the child's summary page, and the child's setting page. As shown in Figure~\ref{fig:web1}, the user can click on an individual child's name in the bottom left corner to enter a child's activity page in Figure~\ref{fig:web2}. Then a user in the page shown in Figure~\ref{fig:web2} can click on the settings icon on the top left to enter a child's setting page shown in Figure ~\ref{fig:web3}. From the setting page, the user can go back to the homepage shown in Figure ~\ref{fig:web1} by clicking on the home button.


\section{Discussion}
In this section, we first present some recommendations for future designs of the Parent Dashboard based on our findings. We then discuss the limitations of our study and future work.

\subsection{Recommendations for Parent Dashboard}
The Alexa Parental Dashboard is a separate website from the Alexa App. And it is very hard to access from the Alexa App. We propose to combine the Alexa App and Dashboard together. A user-friendly guide for parental control settings in the dashboard is also important. Some features in the current Alexa Parental Dashboard are very hard to find and set up. Additionally, a lot of the setting details are hidden in the sub-menu. These features and settings should be made clearer to the users. Furthermore, there are some features that are missing from the current Alexa Parent Dashboard. For example, we see the lack of a bar graph to identify which activity has the most use time and a pie chart to identify which activity has the largest percentage. In our survey, the participants would like to have more types of data representations. Besides, some parents only allowed their kids to use smart speakers when they were around as an alternative method to parental control mode. However, service providers should aim to develop more comprehensive protections for kid users so that parents can trust smart speakers as a tool to entertain their kids and help them grow.

\subsection{Limitations and Future Work}
Our survey was conducted with 140 participants, which might not be large enough for an in-depth understanding of the problems. Moreover, survey data are self-reported data that might not capture all the preferences that users have. Future work could use an interview or in-lab study to further get insights into the users' preferences.


The prototype for the Parent Dashboard we presented in this study is only an initial design of our envisioned Parent Dashboard. We would like to make more revisions to this design in future work. A future user study can be conducted to evaluate the design and the baseline. Besides, in our study, we only focus on Amazon Alexa. However, our results can be further extended to address other platforms such as Google Assistant.

\section{Conclusion}
In this study, we show that the current design and interaction in Amazon's Parent Dashboard can be improved in multiple aspects. We propose some improvements for the Alexa App and Alexa Parent Dashboard. Service providers may only focus on creating more features for the parents. However, parents may not be technically savvy, and it can become complex and time-consuming for them to get familiar with all the functionalities. Thus, it is important to pay attention to user experience.

%

\bibliographystyle{ACM-Reference-Format}
\bibliography{references}


\end{document}